\documentclass[aps,pre,preprint,showpacs,floatfix]{revtex4-1}
\usepackage{bm}
\usepackage{amsmath}
\usepackage{epsfig}
\usepackage{rotating}
\usepackage{graphicx}
\usepackage{rotating}

\def\be{\begin{equation}}
\def\lan{\left\langle}
\def\ran{\right\rangle}
\def\ee{\end{equation}}
\def\barr{\begin{array}}
\def\earr{\end{array}}

\def\l{\left}
\def\r{\right}
\def\dis{\displaystyle}
\def\ed{\end{document}}
\def\f{\frac}

\def\co{{\cal O}}

\def\sbin{\l[\begin{smallmatrix} n \\k \end{smallmatrix}\r]}
\def\ed{\end{document}}
\begin{document}

\title{Bivariate $q$-normal distribution for transition strengths distribution
from many-particle random matrix ensembles generated by  $k$-body interactions}

\author{V. K. B. Kota} 

\affiliation{Physical Research  Laboratory, Ahmedabad 380 009, India}

\email{vkbkota@prl.res.in}

\author{Manan Vyas}

\affiliation{Instituto de Ciencias F{\'i}sicas, Universidad Nacional
Aut{\'o}noma de M\'{e}xico, 62210 Cuernavaca, M\'{e}xico}

\email{manan@icf.unam.mx}

\begin{abstract}

Recently it is established, via lower order moments, that the univariate
q-normal distribution, which is the weight function for $q$-Hermite polynomials,
describes the ensemble averaged eigenvalue density from many-particle random
matrix ensembles generated by $k$-body interactions [Manan Vyas and V.K.B. Kota,
J. Stat. Mech. {\bf 2019}, 103103 (2019)]. These ensembles are generically
called embedded ensembles of $k$-body interactions [EE($k$)] and their GOE and
GUE versions are called EGOE($k$) and EGUE($k$) respectively. Going beyond this
work, the lower order bivariate reduced moments of the transition strength
densities, generated by EGOE($k$) [or EGUE($k$)] for the Hamiltonian and an
independent EGOE($t$) for the transition operator $\co$ that is $t$-body, are
used to establish that the  ensemble averaged bivariate transition densities
follow the bivariate $q$-normal distribution. Presented are also formulas for
the bivariate correlation coefficient $\rho$ and the $q$ values as a function of
the particle number $m$, number of single particle states $N$ that the particles
are occupying and the body ranks $k$ and $t$ of $H$ and $\co$ respectively. 
Finally, using the bivariate $q$ normal form a formula for the chaos measure
number of principal components (NPC) in the transition strengths from a state
with energy $E$ is presented.

\end{abstract}

%\pacs{}

\maketitle

\section{Introduction}

Statistical properties of isolated finite many-particle systems such as atomic
nuclei,  mesoscopic systems (quantum dots, small metallic grains), interacting
spin systems modeling quantum computing core, ultra-cold atoms, quantum black
holes using SYK model and so on are being investigated with  renewed interest in
recent years for deeper understanding of quantum many-body chaos and
thermalization in finite quantum systems. It is now well established that Random
matrix theory    is appropriate for providing answers to many of the questions
in this topic. See Refs. \cite{kota,zel-lea,Rigol,Verb-1,Verb-2} and references
therein. In most of the finite many-particle quantum systems, their constituents
predominantly interact via few-particle interactions. Therefore, modification of
the  classical Gaussian orthogonal (GOE) or unitary (GUE) or symplectic (GSE)
random matrix ensembles with various deformations, incorporating information
about interactions is essential. An appropriate model is to consider $m$
particles (in the present paper we will restrict to fermions) occupying $N$
single particle (sp) states and interacting with a $k$-body ($k < m$)
interaction. In this situation, using a GOE/GUE/GSE representation for the
Hamiltonian in $k$ particle spaces (defining random $k$-body interactions) and
then propagating the information in the interaction to many particle spaces, we
have embedded ensembles of $k$ particle interactions [EE($k$)] in $m$-particle
spaces. Note that in these ensembles, a GOE/GUE/GSE random matrix ensemble in
$k$-particle spaces is embedded  in the $m$-particle $H$ matrix. Then, with GOE
embedding, we have embedded Gaussian orthogonal ensemble of $k$-body
interactions [EGOE($k$)] and similarly with GUE embedding EGUE($k$) \cite{kota}.
The two-body ensembles are first introduced in \cite{Fr-71,Bo-71} with reference
to nuclear shell model and the seminal paper of Mon and French \cite{MF} gave
first analytical results for the general EGOE($k$). These early papers gave the
remarkable result that as $k$ changes from 1 to $m$, EGOE($k$) [similarly
EGUE($k$)] generates Gaussian to semi-circle transition in the eigenvalue
density \cite{Br-81}. A more modern discussion of this results is due to
Weidenm\"{u}ller \cite{BW}.

Most recently, Verbaarschot and collaborators extended the EGOE concept to the
so called SYK model and pointed out that the weight function (giving orthogonal
property) for $q$-Hermite polynomials describes the Gaussian to semi-circle
transition in the eigenvalue density giving a functional for this transition
\cite{Verb-1}. This weight function is called $q$-normal distribution in
\cite{Sza-1} and throughout this paper we will use this name and its explicit
form is given in Section 2. Using these observations combined with the asymptotic
formulas for the lower order moments of the eigenvalue density generated by   
EGOE($k$) and EGUE($k)$ (both for fermion and boson systems), it is shown in a
previous paper \cite{MaKo-19} that the $q$-normal distribution indeed gives the
eigenvalue density for any $k$ in these ensembles and used here are the lower
order moments of $q$-normal given in \cite{Ism-87}. In \cite{MaKo-19}, derived
are also formulas for the parameter $q$ as a function of $(m,N,k)$. This result
is also found to extend to the strength functions (also called local density of
states).

Going beyond the eigenvalue densities, most important quantities in spectroscopy
are transition strengths generated by a transition operator $\co$. Given an
eigenstate $\l|E_i\ran$ of $H$ in a $m$ particle space, action of $\co$ on this
state will result in the transition to states $\l|E_f\ran$ with transition
probability or transition strength $\l|\lan E_f \mid \co \mid E_i\ran\r|^2$.
Multiplying this with the eigenvalue densities at $E_i$ and $E_f$ will give
transition strength densities $\rho_{biv-\co}(E_i,E_f)$. In the situation that
$\co$ a $t$-body operator, representing $H$ and $\co$ by independent EGOE($k$)
and EGOE($t$), it was shown via the lower order moments of
$\rho_{biv-\co}(E_i,E_f)$ that it will take bivariate Gaussian form  for $(k,t)
<< m$ (also assuming the dilute limit with $m \rightarrow \infty$,  $N
\rightarrow \infty$ and $m/N \rightarrow 0$) \cite{FKPT,KoMa-15}. This result is
used in several applications in nuclear structure, for example to calculate
$\beta$-decay rates for pre-super novae stars, nuclear structure matrix elements
for neutrinoless double beta decay and so on \cite{KM-94,KH-17}. An important
unanswered question here is about the form of $\rho_{biv-\co}$ for all $k \leq
m$ and $t \leq m$. The purpose of the present paper is to address this question
and establish that indeed the form of $\rho_{biv-\co}$ in general will be
bivariate $q$-normal distribution giving bivariate normal (Gaussian) form as $q
\rightarrow 1$ and a bivariate semi-circle for $q=0$. Now we will give a
preview.

In Section 2, we will introduce $q$-Hermite polynomials, $q$-normal distribution
and also the bivariate $q$-normal distribution. Also presented here are some of
their important properties. All the results in this Section are from
\cite{Sza-1,Sza-2}. In Section 3, we will derive formulas the reduced bivariate
moments $\mu_{rs}$, $r+s \leq6$ of the bivariate $q$-normal distribution. Using
these and the known results for EGOE and EGUE, in Section 4 established is the
main result that the $\rho_{biv-\co}(E_i,E_f)$ follow bivariate $q$-normal form.
Presented are also formulas for the bivariate correlation coefficient $\rho$
and the $q$ values, that define a bivariate $q$-normal, as a function of
$(m,N,k,t)$. In Section 5, as an application of the bivariate $q$-normal, a
formula in terms of an integral is given for the chaos measure number of
principle components (NPC) in the transition strengths originating from a
initial eigenstate of a $m$ particle Hamiltonian. Finally, Section 6 gives 
conclusions.

\section{$q$-Hermite polynomials and bivariate $q$-normal distribution}

Let us begin with $q$-numbers $\l[n\r]_q$, $q$ factorials $\l[n\r]_q!$ and
$q$-binomials $\sbin_q$,
\be
\barr{l}
\l[n\r]_q = \dis\frac{1-q^n}{1-q} = 1+q +q^2 + \ldots +  q^{n-1}\;;\;\;
\l[0\r]_q=0 \;,\\
\l[n\r]_q! = \dis\prod_{j=1}^{n} \l[j\r]_q!\;; \;\; \l[0\r]_q!=1\;,\\
\sbin_q = \dis\frac{\l[n\r]_q!}{\l[n-k\r]_q!\;\l[k\r]_q!}\;;\;\;n \geq k 
\geq 0\;.
\earr \label{eq.qbiv-1}
\ee
Note that $\l[n\r]_{q=1} = n$, $\l[n\r]_{q=1}! = n!$ and $\sbin_{q=1} =
{n\choose k}$. Although we can use $-1 \leq q \leq +1$, in the applications in
this paper $0 \leq q \leq 1$.  With the $q$ numbers, the $q$-Hermite polynomials
are defined by the relation
\be
H_{n+1}(x|q) = x H_n(x|q) - \l[n\r]_q H_{n-1}(x|q)\;\;\mbox{with}\;\; 
n \geq 1, \;\;H_{-1}(x|q)=0,\;\;H_0(x|q)=1\;.
\label{eq.qbiv-2}
\ee
Note that $H_n(x|1)=H_n(x)$, the Hermite polynomials with respect to 
$1/\sqrt{2\pi}\,\exp-x^2/2$. Also, $H_n(x|0)=U_n(x/2)$, the Chebyshev
polynomials that satisfy the relation
$$
2x U_n(x) = U_{n+1}(x) + U_{n-1}(x)\;;\;\;\;U_{-1}(x)=0,\;U_0(x)=1\;.
$$
Now, let us introduce the $q$-normal distribution $f_{qN}(x|q)$,
\be
f_{qN}(x|q) = \dis\frac{\dis\sqrt{1-q} \dis\prod_{\kappa=0}^{\infty} \l(1-
q^{\kappa +1}\r)}{2\pi\,\dis\sqrt{4-(1-q)x^2}}\; \dis\prod_{k=0}^{\infty} 
\l[(1+q^k)^2 - (1-q) q^k x^2\r]\;.
\label{eq.qbiv-3}
\ee
The $f_{qN}(x|q)$ is defined over $S(q)$ with
$$
S(q) = \l(-\dis\frac{2}{\dis\sqrt{1-q}}\;,\;+\dis\frac{2}{\dis\sqrt{1-q}}\r) 
$$
and $q$ in this work takes values $0$ to $1$. For $q=1$ taking the limit
properly will give $S(q)=(-\infty , \infty)$. Note that the integral of
$f_{qN}(x|q)$ over $S(q)$ is unity. It is easy to see that
$f_{qN}(x|1)=1/\sqrt{2\pi}\,\exp-x^2/2$, the Gaussian and $f_{qN}(x|0)=(1/2\pi)
\sqrt{4-x^2}$, the semi-circle. A very important property of $f_{qN}(x|q)$ is 
that it is the weight function with respect to which the $q$-Hermite polynomials
are orthogonal over $S(q)$ giving,
\be
\dis\int_{S(q)} H_n(x|q) H_m(x|q) f_{qN}(x|q) dx = \l[n\r]_q!\;\delta_{mn}\;.
\label{eq.qbiv-4}
\ee
Going further, bivariate $q$-normal distribution $f_{biv-qN}(x,y|\rho , q)$ as 
given in \cite{Sza-1} is defined as follows,
\be
\barr{l}
f_{biv-qN}(x,y|\rho , q) = f_{qN}(x|q) f_{qN}(y|q) h(x,y|\rho , q)\;;\\
h(x,y|\rho ,q)= \dis\prod_{k=0}^\infty \dis\frac{1-\rho^2 q^k}{
(1-\rho^2 q^{2k})^2 -(1-q)\,\rho\, q^k\,(1+\rho^2 q^{2k})\,xy +
(1-q)\rho^2 q^{2k} (x^2 +y^2)}\;,
\earr \label{eq.qbiv-5}
\ee
where $\rho$ is the bivariate correlation coefficient. The conditional 
$q$-normal densities ($f_{CqN}$) are then,
\be
\barr{l}
f_{biv-qN}(x,y|\rho , q) = f_{qN}(x|q) f_{CqN}(y|x; \rho , q) =
f_{qN}(y|q) f_{CqN}(x|y; \rho , q)\;;\\
f_{CqN}(x|y; \rho , q)=f_{qN}(x|q) h(x,y|\rho ,q)\;,\\
f_{CqN}(y|x; \rho , q)=f_{qN}(y|q) h(x,y|\rho ,q)\;.
\earr \label{eq.qbiv-6}
\ee
A very important property of $f_{CqN}$ is
\be
\dis\int_{S(q)} H_n(x|q) f_{CqN}(x|y; \rho ,q) dx = \rho^n H_n(y|q)\;.
\label{eq.qbiv-7}
\ee
Putting $n=0$ in Eq. (\ref{eq.qbiv-7}), it is easy to infer that $f_{CqN}$ and 
hence $f_{biv-qN}$ are normalized to unity over $S(q)$. We will make use of 
Eqs. (\ref{eq.qbiv-4}) and (\ref{eq.qbiv-7}) in the next Section to arrive at 
the main result of this paper given in Section 4. Let us mention that for 
$q=1$ and $0$, $f_{CqN}$ reduces to 
\be
\barr{l}
f_{CqN}(x|y;\rho , q=1) = \dis\frac{1}{2\pi (\dis\sqrt{1-\rho^2})}\,
\exp -\dis\frac{(x-\rho y)^2}{2(1-\rho^2)}\;,\\
f_{CqN}(x|y;\rho ,q=0) = \dis\frac{(1-\rho^2) \dis\sqrt{4-x^2}}{
2\pi \l[ (1-\rho^2)^2 -\rho(1+\rho^2)xy +\rho^2(x^2+y^2)\r]}\;.
\earr \label{eq.qbiv-8}
\ee
There are many other properties of $q$-Hermite polynomials and $f_{CqN}$ as
given in detail in \cite{Sza-1,Sza-2}. Some of these are,
\be
\barr{l}
f_{biv-qN}(x,y|\rho , q) = f_{qN}(x|q) f_{qN}(y|q) \dis\sum_{n=0}^{\infty}
\dis\frac{\rho^n}{\l[n\r]!} H_n(x|q) H_n(y|q)\;,\\
\phi(x,t|q)= \dis\prod_{k=0}^{\infty} \l(1-(1-q)xtq^k+(1-q)t^2q^{2k}\r)^{-1} 
= \dis\sum_{j=0}^{\infty} \dis\frac{t^j}{\l[j\r]_q!}\,H_j(x|q)\;,\\
\dis\int_{S(q)} P_n(x|y,\rho ,q) P_m(x|y,\rho , q)f_{CqN}(x|y; \rho ,q) dx 
= \l(\rho^2\r)_n \l[n\r]_q!\, \delta_{mn}\;.
\earr \label{eq.qbivhh}
\ee
The first equality here can be used for example to obtain Eq. (\ref{eq.qbiv-7}).
The Second equality gives the generating $\phi(x,t|q)$ of the $q$-Hermite
polynomials. In the third equality, $P_n(x|y,\rho, q)$ are Al-Salam-Chihara
polynomials and $(\rho^2)_n=\dis\prod_{i=0}^{n-1} (1-\rho^2 q^i)$ with
$(\rho^2)_0=1$. Now, we will  derive formulas for the reduced bivariate moments
$\mu_{rs}$ of $f_{biv-qN}$.  

\section{Reduced bivariate moments $\mu_{r+s}$, $r+s \leq 6$ of bivariate 
$q$-normal}

Reduced bivariate central moments $\mu_{rs}$ of $f_{biv-qN}$ are defined by
\be
\mu_{rs} = \dis\int_{S(q)} x^r\,y^s\, f_{biv-qN}(x,y|\rho ,q)\, dx dy\;.
\label{eq.qbiv-9}
\ee
As $H_0(x|q)=1$, $x=H_1(x|q)$ and $x^2=H_2(x|q)+1$, using Eqs. (\ref{eq.qbiv-4})
and (\ref{eq.qbiv-7}) will immediately give (note that the integrals of $f_{qN}$
and $f_{biv-qN}$ are 1) the results $\mu_{10}=\mu_{01}=0$ and
$\mu_{20}=\mu_{02}=1$. Also $\mu_{rs}=\mu_{sr}$ and $\mu_{rs}=0$ for $r+s$ odd.
As lower order moments suffice to arrive at the ensemble averaged forms of
$\rho_{biv-\co}$, here we  will consider only $\mu_{rs}$ of $f_{biv-qN}$ with
$r+s=4$ and $6$  and $r \geq s$. To derive the formulas for $\mu_{rs}$, we will
first write $x^p$, $p \leq 6$ in terms of $H_n(x|q)$, $n \leq 6$ using Eq.
(\ref{eq.qbiv-2}). This will give, after some algebra the formulas,
\be
\barr{rcl}
x & = & H_1\;,\;\;\;x^2 =  H_2+H_0\;,\\
x^3 & = & H_3 + (2+q)H_1\;,\\
x^4 & = & H_4 + (3+2q+q^2)H_2 +(2+q)H_0\;,\\
x^5 & = & H_5 +(4+3q+2q^2+q^3)H_3 +(5+6q+3q^2+q^3)H_1\;,\\
x^6 & = & H_6 + (5+4q+3q^2+2q^3+q^4)H_4 \\
& & + (9+13q+12q^2+7q^3+3q^4+q^5)H_2 +(5+6q+3q^2+q^3)H_0\;.
\earr \label{eq.qbiv10}
\ee
Here, $H_n$ stands for $H_n(x|q)$ and $H_0(x|q)=1$. Firstly, it is easy to see 
that $\mu_{11}=\rho$,
\be
\barr{rcl}
\mu_{11} & = & \dis\int_{S(q)} x\,y\, f_{biv-qN}(x,y|\rho ,q)\, dx dy = 
\dis\int_{S(q)} x f_{qN}(x|q) dx \dis\int_{S(q)} H_1(y) f_{CqN}(y|x;\rho ,q) 
dy \\
& = & \rho \;\dis\int_{S(q)} H_1(x) H_1(x) f_{qN}(x|q) dx=\rho\;.
\earr \label{eq.qbiv11}
\ee
In the first step here we have used Eqs. (\ref{eq.qbiv10}) and
(\ref{eq.qbiv-7}) and
in the second step Eq. (\ref{eq.qbiv-4}). For $r+s=4$ we need $\mu_{40}$, 
$\mu_{31}$ and $\mu_{22}$. The $\mu_{40}$ is simple,
\be
\mu_{40} = \dis\int_{S(q)} x^4 f_{N}(x|q) dx = (2+q) \dis\int_{S(q)} 
f_{N}(x|q) dx = (2+q)\;.
\label{eq.qbiv-12}
\ee 
In the above we have substituted for $x^4$ the expansion in terms of $H_n$ using
Eq. (\ref{eq.qbiv10})  and then used Eq. (\ref{eq.qbiv-4}). Similarly, formula
for $\mu_{31}$ is,
\be
\barr{rcl}
\mu_{31} & = & \dis\int_{S(q)} x^3\,y\, f_{biv-qN}(x,y|\rho ,q)\, dx dy = 
\dis\int_{S(q)} x^3 f_{qN}(x|q) dx \dis\int_{S(q)} H_1(y) 
f_{CqN}(y|x;\rho ,q) dy \\
& = & \rho\;\dis\int_{S(q)} \l[H_3(x|q) H_1(x|q) + (2+q)H_1(x|q) H_1(x|q)\r] 
f_{qN}(x|q) dx = 
\rho (2+q) = \rho \mu_{40}\;.
\earr \label{eq.qbiv-12a}
\ee
Finally, proceeding to $\mu_{22}$ we have,
\be
\barr{rcl}
\mu_{22} & = & \dis\int_{S(q)} x^2\,y^2\, f_{biv-qN}(x,y|\rho ,q)\, dx dy = 
\dis\int_{S(q)} x^2 f_{qN}(x|q) dx \dis\int_{S(q)} \l[H_2(y|q)+1\r] 
f_{CqN}(y|x;\rho ,q) dy \\
& = & \dis\int_{S(q)} \l[H_2(x|q) +1\r] \l[\rho^2 H_2(x|q) +1\r] 
f_{qN}(x|q) dx = 1 + (1+q)\rho^2\;.
\earr \label{eq.qbiv-13}
\ee   
Turning to the sixth order moments first we have easily using $x^6$ and $x^5$ 
from Eq. (\ref{eq.qbiv10}),
\be
\mu_{60} = (5+6q+3q^2+q^3)\;,\;\;\;\mu_{51}= \rho \mu_{60}\;.
\label{eq.qbiv14}
\ee
Formula for $\mu_{42}$ is,
\be
\barr{rcl}
\mu_{42} & = & \dis\int_{S(q)} x^4\,y^2\, f_{biv-qN}(x,y|\rho ,q)\, dx dy \\
& = & 
\dis\int_{S(q)} x^4 f_{qN}(x|q) dx \dis\int_{S(q)} \l[H_2(y|q)+1\r] 
f_{CqN}(y|x;\rho ,q) dy \\
& = & \dis\int_{S(q)} \l[H_4(x|q) + (3+2q+q^2)H_2(x|q)+(2+q)\r] 
\l[\rho^2 H_2(x|q) +1\r] f_{qN}(x|q) dx \\
& = & \rho^2 (3+2q+q^2) \l[2\r]_q! + (2+q) = (2+q) + \rho^2(3+5q+3q^2+q^3)\;.
\earr \label{eq.qbiv15}
\ee
Finally, $\mu_{33}$ is given by
\be
\barr{rcl}
\mu_{33} & = & \dis\int_{S(q)} x^3\,y^3\, f_{biv-qN}(x,y|\rho ,q)\, dx dy \\
& = & 
\dis\int_{S(q)} x^3 f_{qN}(x|q) dx \dis\int_{S(q)} \l[H_3(y|q)+(2+q)H_1(y|q)\r] 
f_{CqN}(y|x;\rho ,q) dy \\
& = & \dis\int_{S(q)} \l[H_3(x|q)+(2+q)H_1(x|q)\r] \l[\rho^3 H_3(x|q) +
\rho (2+q) H_1(x|q)\r] f_{qN}(x|q) dx \\
& = & (2+q)^2 \rho + (1+q)(1+q+q^2)\rho^3 \;.
\earr \label{eq.qbiv16}
\ee
Formulas for the bivariate moments given in Eqs. (\ref{eq.qbiv-12}) -
(\ref{eq.qbiv16}) can be derived also from the formulation presented in
\cite{Sza-3}.  Now, we will consider the bivariate moments of the transition
strength densities generated by EGOE (and EGUE) and establish that the strength
densities follow $f_{biv-qN}$ form.

\begin{table}
\caption{Reduced bivariate moments $\mu_{rs}$ from EGOE for a system of $m=10$
fermions with $k=2-6$ and $t=1$ and $2$. The results follow from Eqs. (20)-(23).
Given also are values of the bivariate correlation coefficient $\rho$ and the $q$
values. Numbers in the brackets give the difference between EGOE values and
those from the bivariate $q$-normal.  Note that, for $k=6$ or higher the
corrections to the $\mu_{rs}$ are $0$ and therefore the results for $k \geq 7$
are not shown in the table.}
\begin{tabular}{c|c}
\hline
\hline
$k=2,t=1$ & $k=2,t=2$ \\
$\rho = 0.8$,$\;\;q= 0.622$ & $\rho=0.622$,$\;\;q=0.622$ \\
$\mu_{22}= 2.013 +(-0.025)$,$\;\;\mu_{60}= 10.102 +(-0.034)$  & 
$\mu_{22}=1.595 +(-0.034)$, $\;\;\mu_{60}=10.102 +(-0.034)$ \\ 
$\mu_{42}= 7.095 +(-0.336)$, $\;\;\mu_{33}= 7.042 +(-0.128)$ & 
$\mu_{42}=5.246+(-0.285)$, $\;\;\mu_{33}=4.943+ (-0.12)$ \\
$k=3,t=1$ & $k=3,t=2$ \\
$\rho = 0.7$,$\;\;q=0.292$ & $\rho=0.467$,$\;\;q=0.292$ \\ 
$\mu_{22}=1.607 +(-0.026)$,$\;\;\mu_{60}=7.015 + (-0.015) $ &
$\mu_{22}=1.257+(-0.025)$,$\;\;\mu_{60}=7.015+(-0.015)$ \\
$\mu_{42}=4.47 +(-0.144)$,$\;\;\mu_{33}=4.222 +(-0.064)$ &
$\mu_{42}=3.213+(-0.111)$, $\;\;\mu_{33}= 2.595+(-0.036)$ \\
$k=4,t= 1$ & $k=4,t=2$ \\
$\rho= 0.6$,$\;\;q=0.071$ & $\rho=0.333$,$\;\;q=0.071$ \\
$\mu_{22}= 1.374+(-0.011)$, $\;\;\mu_{60}= 5.444 +(0.0)$ & 
$\mu_{22}=1.113+(-0.006)$,$\;\;\mu_{60}=5.444+(0.0)$\\
$\mu_{42}=3.248+(-0.038)$,$\;\;\mu_{33}=2.808+(-0.015)$ & 
$\mu_{42}= 2.426+(-0.021)$,$\;\;\mu_{33}=1.468+(-0.005)$ \\
$k=5,t= 1$ & $k=5,t=2$ \\
$\rho= 0.5$,$\;\;q= 0.004$ & $\rho=0.222$,$\;\;q=0.004$ \\
$\mu_{22}=1.25+(-0.001)$,$\;\;\mu_{60}=5.024+(0.0)$ & 
$\mu_{22}=1.049+(0.0)$,$\;\;\mu_{60}=5.024+(0.0)$   \\
$\mu_{42}=2.756+(-0.003)$, $\;\;\mu_{33}= 2.133 +(-0.001)$ & 
$\mu_{42}=2.153+(-0.001)$,$\;\;\mu_{33}=0.903+(0.0)$ \\
$k=6,t=1$ & $k=6,t=2$ \\
$\rho=0.4$,$\;\;q=0.0$ & $\rho=0.133$,$\;\;q=0.0$ \\
$\mu_{22}=1.16+(0.0)$,$\;\;\mu_{60}=5+(0.0)$ & $\mu_{22}=1.018+(0.0)$, 
$\;\;\mu_{60}=5 +(0.0)$ \\
$\mu_{42}=2.48+(0.0)$, $\;\;\mu_{33}= 1.664+(0.0)$ & 
$\mu_{42}=2.053+(0.0)$, $\;\;\mu_{33}=0.536+(0.0)$ \\
\hline
\hline
\end{tabular}
\label{tab1}
\end{table}
 
\section{Bivariate $q$-normal representing bivariate transition strength
densities generated by EGOE and EGUE}

Let us say we have a system of $m$ fermions occupying $N$ number of sp states
and the Hamiltonian ($H$) operator is $k$-body. Then, the $m$ particle space
dimension is ${N \choose m}$. Starting with $H(k)$, it is possible to construct
the $m$ particle $H$ matrix and obtain the eigenstates $\l|E\ran$ with energy
$E$ in $m$ particle spaces. Now, given a $t$-body transition operator $\co(t)$
acting on an eigenstate $\l|E_i\ran$ in the $m$ particle space will populate the
$m$ particle state  $\l|E_f\ran$ with probability $\l|\lan E_f \mid \co \mid
E_i\ran\r|^2$ and the resulting bivariate transition strength density
(normalized to unity) is,
\be
\rho_{biv-\co}(E_i,E_f) = \l[\lan \lan \co^\dagger \co \ran\ran^m\r]^{-1}\; 
\lan \lan \co^\dagger \delta(H-E_f) \co \delta(H-E_i)\ran \ran^m\;.
\label{eq.qbiv17}
\ee
Note that $\lan \lan X \ran \ran^m =\sum_E \lan E \mid X \mid E\ran$ where 
$\l|E\ran$ are all the eigenstates of the $m$ particle Hamiltonian matrix.  In
order to derive the statistical law for the form of $\rho_{biv-\co}(E_i,E_f)$,
random matrix theory is used by representing the $H$ by EGOE($k$) and the $\co$
by an independent EGOE($t$). With this, formulas for the (ensemble averaged)
bivariate reduced central moments $\mu_{rs}$ of $\rho_{biv-\co}(E_i,E_f)$ are
derived, as a function of $(m,k,t)$ using the so called binary correlation
approximation for $r+s=4$ and $6$ (also for $\mu_{11}$); 
see Refs. \cite{FKPT,Ko-01}.
These results are also valid for the EGUE($k$) for $H$ and EGUE($t$) for $\co$;
see \cite{kota}. Further, for $\mu_{11}$ and $\mu_{rs}$ with $r+s=4$ results with
finite $N$ corrections are  derived in \cite{KoMa-15}. Quite strikingly, the
formulas are close to those obtained for $f_{biv-qN}$. We will describe this in
some detail below starting with the formulas without finite $N$ corrections.

\subsection{Equivalence between lower order moments}

With EGOE($k$) for $H$ and EGOE($t$) for $\co$, the bivariate reduced  central
moments $\mu^{E}_{rs}$ for $r=s=1$ (the superscript $E$ denoting that the
quantities are for the EGOE ensemble) and for $r+s=4$, using binary correlation
approximation and the dilute limit conditions with $N \rightarrow \infty$ as
described in \cite{FKPT,Ko-01,MaKo-19}, are given by
\be
\barr{rcl}
\mu^{E}_{11} & = & \dis\frac{{m-t \choose k}}{{m \choose k}} \Rightarrow 
\rho^E=\dis\frac{{m-t \choose k}}{{m \choose k}}\;,\\
\mu^E_{40}=\mu^E_{04} & = & 2 + \dis\frac{{m-k \choose k}}{{m \choose k}} 
\Rightarrow q^E = \dis\frac{{m-k \choose k}}{{m \choose k}}\;,\\
\mu^E_{31}=\mu^E_{13} & = & \rho^E \mu_{40}\;,\\
\mu^E_{22} & = & 1 + \l(\rho^E\r)^2 (1+q^E)+ \rho^E \Delta_0\;;\\  
\Delta_0 & = & \dis\frac{{m-k-t \choose k}}{{m \choose k}} -
\dis\frac{{m-t \choose k} {m-k \choose k}}{{m \choose k}^2}\;.
\earr \label{eq.qbiv18}   
\ee
Thus, $\mu_{11}$ gives the EGOE formula for the bivariate correlation
coefficient $\rho^E$ and $\mu^E_{40}$ gives the formula for the $q^E$ parameter
(see also \cite{MaKo-19}). In terms of these, the formulas for $\mu^E_{31}$ and
$\mu^E_{22}$ given in \cite{FKPT,Ko-01} are rewritten in Eq. (\ref{eq.qbiv18}).
To the extent that the correction $|\rho^E \Delta_0| \sim 0$, the $\mu^E_{rs}$
with $r+s=4$ from EGOE are same as the $\mu_{rs}$ from $f_{biv-qN}$. Numerical
calculations using some typical values for $(m,k,t)$ show that this is indeed
the situation; see Tables \ref{tab1} and \ref{tab2}. Thus, the fourth order EGOE
moments show that $f_{biv-qN}$ is a good representation of $\rho_{biv-\co}$. For
further confirming this important result, we will turn to the sixth order
bivariate moments.  

Firstly, rewriting the formula for $\mu^E_{60}=\mu^E_{06}$ given in
\cite{FKPT,Ko-01} in terms of $q^E$ we have
\be
\barr{l}
\mu^E_{60} = \mu^E_{06} = 5 + 6q^E + 3\l[q^E\r]^2 + \l[q^E\r]^3 + 
q^E \Delta_1\;;\;\;\Delta_1=\dis\frac{{m-2k \choose k}}{{m \choose k}} -
\l[\dis\frac{{m-k \choose k}}{{m \choose k}}\r]^2\;,\\
\mu^E_{51} = \mu^E_{15} = \rho^E\,\mu^E_{60}\;.
\earr \label{eq.qbiv19}
\ee
This is same as Eq. (\ref{eq.qbiv14}) provided the correction $\l|q^E
\Delta_1\r| \sim 0$. Examples in Tables \ref{tab1} and \ref{tab2} confirm that
this correction is indeed small. Using the expressions for $\Delta_0$ and
$\Delta_1$ given in Eqs. (\ref{eq.qbiv18}) and (\ref{eq.qbiv19}), the formula 
for $\mu^E_{42}$ is
\be
\barr{l}
\mu^E_{42} = \mu^E_{24} = (2+q^E) + 3 \l[\rho^E\r]^2 + 5\l[\rho^E\r]^2 q^E
+ 3 \l[\rho^E q^E\r]^2 + \l[\rho^E\r]^2 \l[q^E\r]^3 + \rho^E (X)\;;\\
X= \Delta_0 \l[3 + 2 q^E + \Delta_1 + (q^E)^2\r] + \rho^E q^E \Delta_1 -
\rho^E (q^E)^2 + Y\;,\\
Y=\dis\sum_{\nu=k}^{2k} \dis\frac{{m-2\nu \choose k}{m-t-k \choose \nu -k}
{k \choose 2k-\nu}}{{m \choose k}^2}\;.
\earr \label{eq.qbiv20}
\ee
Similarly, simplifying the formula for $\mu^E_{33}$ we have,
\be
\barr{l}
\mu^E_{33} = \rho^E\l[4 + 4q^E + (q^E)^2\r] + \l\{\rho^E\r\}^3 \l[ 1 + 2q^E 
+2(q^E)^2 + (q^E)^3\r] + \rho^E (Z)\;;\\
Z = 2\Delta_0^2 + 4\rho^E q^E \Delta_0 + 2\rho^E \Delta_0 + \Delta_0\l[ 
(q^E)^2\rho^E + q^E \Delta_0 + \Delta_2\r] + \rho^E q^E (q^E \Delta_0 + 
\Delta_2)\;,\\
\Delta_2 = \dis\frac{{m-t-2k \choose k}}{{m \choose k}} -
\dis\frac{{m-k-t \choose k}}{{m \choose k}} \dis\frac{{m-k \choose k}}
{{m \choose k}}\;.
\earr \label{eq.qbiv21}
\ee
The formulas for $\mu^E_{42}$ and $\mu^E_{33}$ will be same as those from
$f_{biv-qN}$ to the extent that the corrections $\l|\rho^E X\r| \sim 0$ in Eq.
(\ref{eq.qbiv20}) and $\l|\rho^E Z\r| \sim 0$ in Eq. (\ref{eq.qbiv21}). This is
indeed the situation as shown using two examples in Tables \ref{tab1} and
\ref{tab2}.

Results in Tables \ref{tab1} and \ref{tab2} clearly establish that in general
the corrections $\rho^E \Delta_0$, $q^E \Delta_1$, $\rho^E X$ and $\rho^E Z$ for
$\mu_{22}$, $\mu_{60}$, $\mu_{42}$ and $\mu_{33}$, with formulas for these given
in Eqs. (20), (21), (22) and (23) respectively, are indeed less than 2-3\% (in a
few cases they are $\sim 5$\%). Therefore, we conclude that the transition
strength density generated by EGOE (similarly, EGUE) is well represented by the
bivariate $q$-normal distribution. Let us mention that it is well known in
statistics \cite{Kendall} and in random matrix theory \cite{Br-81,KH-10} that
lower order moments generate the form of a probability distribution.

\begin{table}
\caption{Reduced bivariate moments $\mu_{rs}$ from EGOE
for a system of $m=15$ fermions with $k=2-8$ and $t=1$ and $2$. The results 
follow from Eqs. (20)-(23). Given also are values of the bivariate correlation 
coefficient $\rho$ and the $q$ values. Numbers in the brackets give the 
difference between EGOE values and those from the bivariate $q$-normal. Note 
that, for $k=8$ or higher the corrections to the $\mu_{rs}$ are $0$ and 
therefore the results for $k \geq 9$ are not shown in the table.}
{\scriptsize
\begin{tabular}{c|c}
\hline 
\hline  
$k=2,t=1$ & $k=2,t=2$ \\
$\rho= 0.867$, $\;\;q= 0.743$ & $\rho=0.743$, $\;\;q= 0.743$ \\
$\mu_{22}= 2.296+(-0.013)$, $\;\;\mu_{60}= 11.502+(-0.021)$ &
$\mu_{22}= 1.941+(-0.021)$, $\;\;\mu_{60}= 11.502+(-0.021)$ \\
$\mu_{42}= 9.022+(-0.315)$, $\;\;\mu_{33}= 9.034+(-0.09)$ &
$\mu_{42}= 7.302+(-0.285)$, $\;\;\mu_{33}= 7.118+(-0.11)$ \\
$k=3,t=1$ & $k=3,t=2$ \\
$\rho=0.8$, $\;\;q= 0.484$ & $\rho= 0.629$, $\;\;q= 0.484$ \\
$\mu_{22}= 1.93+(-0.019)$, $\;\;\mu_{60}= 8.692+(-0.024)$ &
$\mu_{22}= 1.561+(-0.025)$, $\;\;\mu_{60}= 8.692+(-0.024)$ \\ 
$\mu_{42}= 6.239+(-0.233)$, $\;\;\mu_{33}= 6.157+(-0.082)$ &
$\mu_{42}= 4.747+(-0.199)$, $\;\mu_{33}= 4.434+(-0.076)$ \\
$k=4,t=1$ &$k=4,t=2$ \\ 
$\rho= 0.733$, $\;\;q= 0.242$ & $\rho= 0.524$, $\;\;q= 0.242$ \\
$\mu_{22}= 1.651+(-0.017)$, $\;\;\mu_{60}= 6.632+(-0.008)$ &
$\mu_{22}= 1.323+(-0.018)$, $\;\;\mu_{60}= 6.632+(-0.008)$ \\ 
$\mu_{42}= 4.511+(-0.096)$, $\;\;\mu_{33}= 4.281+(-0.041)$ &
$\mu_{42}= 3.367+(-0.081)$, $\;\;\mu_{33}= 2.836+(-0.029)$ \\
$k=5,t=1$ & $k=5,t=2$ \\
$\rho= 0.667$, $\;\;q= 0.084$ & $\rho= 0.429$, $\;\;q= 0.084$ \\
$\mu_{22}= 1.472+(-0.009)$, $\;\;\mu_{60}= 5.525+(-0.001)$ &
$\mu_{22}= 1.192+(-0.007)$, $\;\;\mu_{60}= 5.525+(-0.001)$ \\  
$\mu_{42}= 3.58+(-0.033)$, $\;\;\mu_{33}= 3.231+(-0.014)$ & 
$\mu_{42}= 2.691+(-0.025)$, $\;\;\mu_{33}= 1.947+(-0.007)$ \\
$k=6,t=1$ & $k=6,t=2$ \\
$\rho= 0.6$, $\;\;q= 0.017$ & $\rho= 0.343$, $\;\;q= 0.017$ \\ 
$\mu_{22}= 1.363+(-0.003)$, $\;\;\mu_{60}= 5.102+(0.0)$ &
$\mu_{22}= 1.118+(-0.002)$, $\;\;\mu_{60}= 5.102+(0.0)$ \\ 
$\mu_{42}= 3.119+(-0.008)$, $\;\;\mu_{33}= 2.661+(-0.003)$ & 
$\mu_{42}= 2.375+(-0.005)$, $\;\;\mu_{33}= 1.435+(-0.001)$ \\
$k=7,t=1$ & $k=7,t=2$ \\
$\rho= 0.533$, $\;\;q= 0.001$ & $\rho= 0.267$, $\;\;q= 0.001$ \\ 
$\mu_{22}= 1.285+(0.0)$, $\;\;\mu_{60}= 5.007+(0.0)$ &  
$\mu_{22}= 1.071+(0.0)$,  $\;\;\mu_{60}= 5.007+(0.0)$ \\   
$\mu_{42}= 2.856+(-0.001)$, $\;\;\mu_{33}= 2.288+(0.0)$ &  
$\mu_{42}= 2.215+(0.001)$, $\;\;\mu_{33}= 1.087+(0.0)$ \\
$k=8,t=1$ & $k=8,t=2$ \\
$\rho= 0.467$, $\;\;q= 0.0$ & $\rho=0.2$, $\;\;q=0.0$ \\
$\mu_{22}= 1.218+(0.0)$, $\;\;\mu_{60}= 5+(0.0)$ &
$\mu_{22}= 1.04+(0.0)$, $\;\;\mu_{60}= 5+(0.0)$ \\ 
$\mu_{42}= 2.653+(0.0)$,  $\;\;\mu_{33}= 1.968+(0.0)$ &  
$\mu_{42}= 2.12+(0.0)$, $\;\;\mu_{33}= 0.808+(0.0)$ \\
\hline 
\hline
\end{tabular}
\label{tab2}}
\end{table}
 
\subsection{Formulas for correlation coefficient $\rho^E$ and parameter $q^E$
with finite $N$ corrections}

\begin{figure}
     \centering
         \includegraphics[width=\textwidth,height=\textwidth]{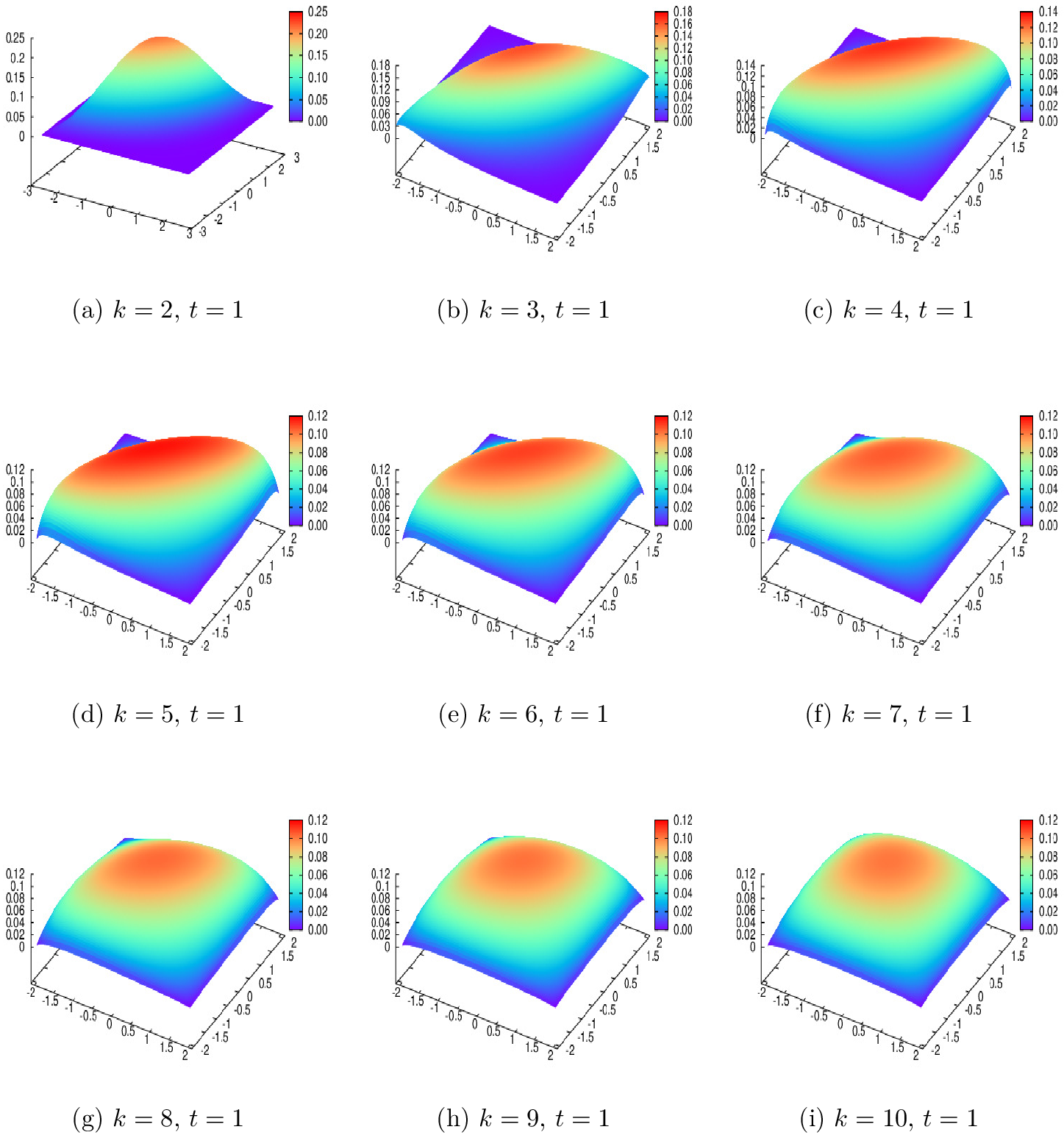}
         \caption{Bivariate transition strength density $f_{biv-qN}(x,y|
	 \rho,q)$ given by Eq. \eqref{eq.qbiv-5} for $m = 10$ fermions in 
	 $N = 20$ sp levels using Eqs. \eqref{eq.qbiv22} and \eqref{eq.qbiv23} 
	 for the parameters $q$ and $\rho$ respectively. Parameters $k$ and $t$ 
	 are as indicated in the figure.}
        \label{fig:bivqN-t1}
\end{figure}

\begin{figure}
     \centering
          \includegraphics[width=\textwidth,height=\textwidth]{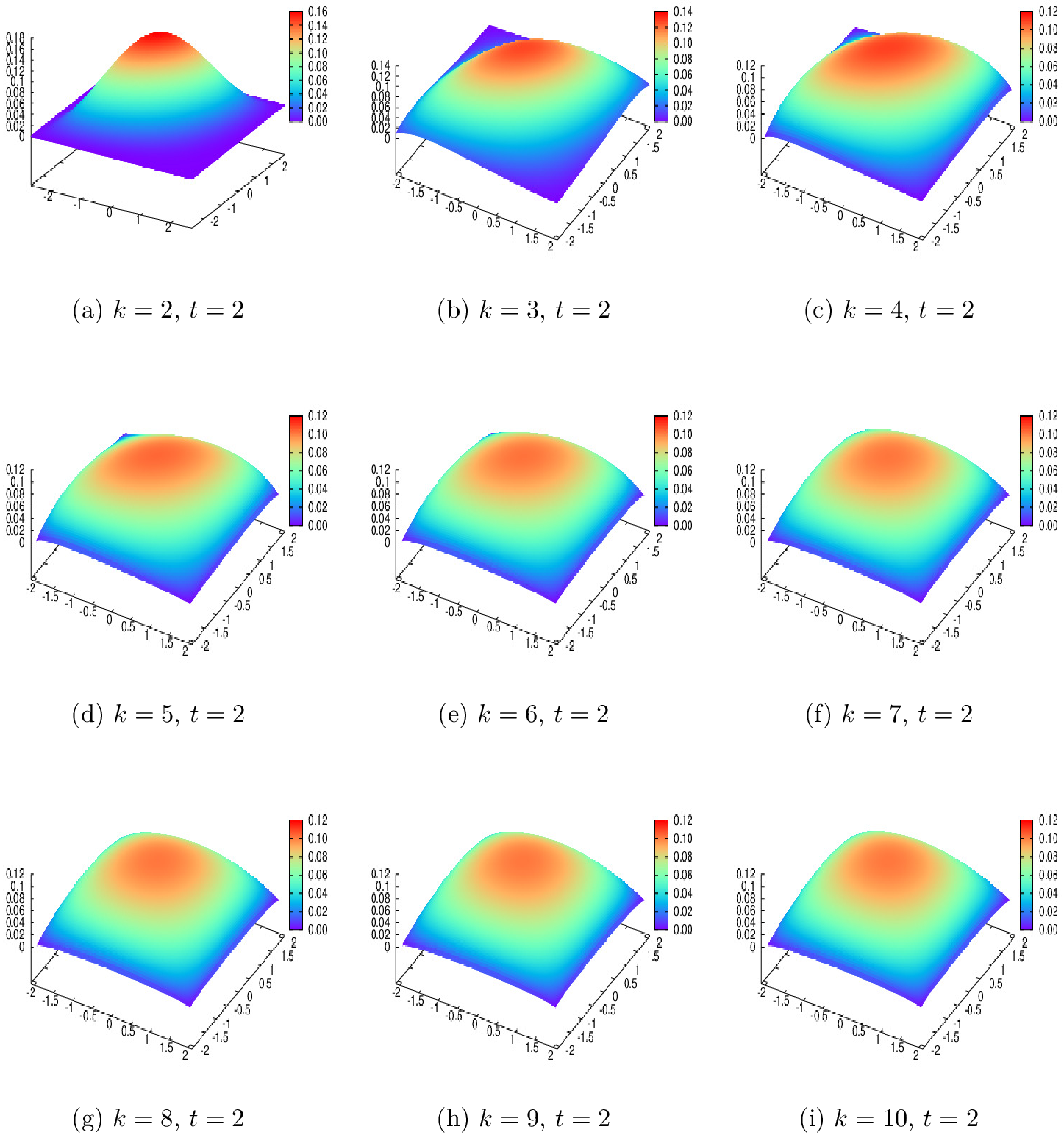}
        \caption{Bivariate transition strength density $f_{biv-qN}(x,y|\rho,q)$
	 given by Eq. \eqref{eq.qbiv-5} for $m = 10$ fermions in $N = 20$ sp 
	 levels using Eqs. \eqref{eq.qbiv22} and \eqref{eq.qbiv23} for the
	 parameters
	 $q$ and $\rho$ respectively. Parameters $k$ and $t$ are as indicated in
	 the figure.}
        \label{fig:bivqN-t2}
\end{figure}

Although in the previous subsection we have used the dilute limit formulas (hence
$N$, the number of sp states do not appear in the formulas), in applying the
bivariate $q$-normal form for the transition strength densities, it is useful to
have formulas for the two parameters $q^E$ and $\rho^E$ with finite $N$
corrections. As it is clearly established earlier in \cite{MaKo-19}, the EGOE
and EGUE give essentially same numerical results for the lower order moments
generating the same form the state densities (similarly for transition strength
densities), we can use Eqs. (13) and (24) given in \cite{KoMa-15}, to write the
formulas for $\rho^E$ and $q^E$ with finite $N$ corrections. For example, the
formula for $q^E$ is, with EGUE($k$) [or EGOE($k$)] representing $H$,
\be
\barr{l}
q^E = \dis\binom{N}{m}^{-1} \dis\sum_{\nu=0}^{min(k,m-k)}\;
\dis\frac{\Lambda^\nu(N,m,m-k)\;\Lambda^\nu(N,m,k)\;d(g_\nu)}{
\l[\Lambda^0(N,m,k)\r]^2} \;;\\ 
\\
\Lambda^\nu(N,m,r) =  \dis\binom{m-\nu}{r}\;\dis\binom{N-m+r-\nu}{r}\;,\\ \\
d(g_\nu)  = \dis\binom{N}{\nu}^2-\dis\binom{N}{\nu-1}^2\;.
\earr \label{eq.qbiv22}
\ee
Note that we are considering $m$ fermions in $N$ sp states with $H$ a $k$-body
operator. Similarly, with $\co$ a $t$-body operator represented by an
independent EGUE($t$) [or EGOE( $t$)], the bivariate correlation coefficient
$\rho^E$ is given by,
\be
\rho^E = \dis\sum_{\nu=0}^{min(t,m-k)}\; 
\dis\frac{\Lambda^\nu(N,m,m-t)\;\Lambda^\nu(N,m,k)\;d(g_\nu)}{\binom{N}{m}\;
\Lambda^0(N,m,k)\;\Lambda^0(N,m,t)} \,.
\label{eq.qbiv23}
\ee
Although we have restricted to $\co(t)$ type operators in this paper, it
is also possible to analyze $\mu^E_{rs}$ with $r+s=4$ and $(rs)=(11)$ for beta
and neutrinoless double beta decay type operators and also for particle removal
operators using the results in \cite{KoMa-15}. More importantly, they will give
formulas, with finite $N$ corrections, for $\rho^E$ and $q^E$ for the transition
strength densities generated by these operators. 

Figure \ref{fig:bivqN-t1} shows the bivariate transition strength density
$f_{biv- qN}(x,y|\rho,q)$ given by Eq. \eqref{eq.qbiv-5} for $m = 10$ fermions
in $N =  20$ sp levels. Parameters $q$ and $\rho$ are calculated using Eqs. 
\eqref{eq.qbiv22} and \eqref{eq.qbiv23} respectively; see Table \ref{tab3} for 
numerical values. Here, $t = 1$ and $k$ varies from 2 to 10. As can be seen
from  this figure, the bivariate transition strength density is close to 
Gaussian form for small $k$ and  becomes semi-circular like with increasing $k$.
Similarly, Figure
\ref{fig:bivqN-t2}  shows the bivariate transition strength density
$f_{biv-qN}(x,y|\rho,q)$ with $t =  2$. The transition in
$f_{biv-qN}(x,y|\rho,q)$ from Gaussian to semi-circular form is faster for 
$t = 2$ in comparison to that for $t = 1$.

\begin{table}
\caption{Correlation coefficient $\rho$ and parameter $q$ with finite-$N$ 
corrections. Values are given for a system of $m = 10$ fermions in $N = 20$ sp 
levels.}
\begin{tabular}{|c|c|c|c|c|c|c|c|}
\hline
\hline
$t$ & $k$ & $\rho$ & $q$ & $t$ & $k$ & $\rho$ & $q$\\
\hline
\hline
 1 & 2 & 0.682 & 0.465 & 2 & 2 & 0.465 & 0.465 \\
  & 3 & 0.559 & 0.176 &  & 3 & 0.314 & 0.176 \\
  & 4 & 0.455 & 0.044 &  & 4 & 0.210 & 0.044 \\
  & 5 & 0.364 & 0.007 &  & 5 & 0.136 & 0.007 \\
  & 6 & 0.284 & 0.001 &  & 6 & 0.085 & 0.001 \\
  & 7 & 0.214 & 0.000 &  & 7 & 0.050 & 0.000 \\
  & 8 & 0.152 & 0.000 &  & 8 & 0.027 & 0.000 \\
  & 9 & 0.096 &  0.000 &  & 9 & 0.012 & 0.000 \\
  & 10 & 0.046 & 0.000 &  & 10 & 0.004 & 0.000 \\
\hline
\hline
\end{tabular}
\label{tab3}
\end{table}

\section{Application of bivariate $q$-normal form of the strength densities}

Using the bivariate $q$-normal form for the strength densities and using the
formulation given in \cite{KS-PLB,KGK-PRE}, it is possible to derive formulas
for the chaos measures number of principle components (NPC) and information
entropy in transition strengths. For example, (NPC)$_E$ in transition strengths
generated by the action of a transition operator $\co(t)$ on an eigenstate with
energy $E$ [of a given $(m,N)$ system with $k$-body interactions] gives the 
number of $m$-particle eigenstates excited by the transition operator. Note that,
(NPC)$_E$ is small implies that the state $E$ is collective or regular with
respect to $\co$ and if it is large then the state is chaotic or mixed. Eq. (6)
of \cite{KS-PLB} gives,
\be
(\mbox{NPC})_E = \dis\frac{d}{3}\;\l[\rho_{1:\co}(E)\r]^2\;
\l\{\dis\int_{\epsilon_0^i}^{\epsilon_0^f} dE_f\;\dis\frac{\l[
\rho_{biv-\co}(E,E_f)\r]^2}{\rho(E_f)}\r\}^{-1}\;.
\label{eq.npc1}
\ee
Here, $d=\binom{N}{m}$ is the dimension of the space, $\rho(E_f)$ is the
normalized state density of the final states with energy $E_f$, $\rho_{biv-\co}$
is the normalized bivariate transition strength density and $\rho_{1:\co}$ is the
marginal density of $\rho_{biv-\co}$. We will give the values of $\epsilon_0^i$
and $\epsilon_0^f$ ahead. Putting the centroids and widths
$(\epsilon_1,\sigma_1)$ and $(\epsilon_2,\sigma_2)$ of $E$ and $E_f$
respectively in $\rho_{biv-\co}$ and similarly the centroid and width
$(\epsilon_f, \sigma_f)$ of $\rho(E_f)$ we have from Sections II and IV,
\be
\barr{l}
\rho(E_f) = \dis\frac{1}{\sigma_f}\;f_{qN}\l(\f{E-\epsilon_f}{\sigma_f} \mid 
q^\prime\r) \\
\mbox{with}\;\;\epsilon_f - \dis\frac{2\sigma_f}{\dis\sqrt{1-q^\prime}} 
\leq E_f \leq \epsilon_f + \dis\frac{2\sigma_f}{\dis\sqrt{1-q^\prime}} \;,\\
\\
\rho_{biv-\co}(E,E_f)=\dis\frac{1}{\sigma_1 \sigma_2}\;f_{biv-qN} \l(\hat{E},
\f{E_f-\epsilon_2}{\sigma_2}\mid \rho , q\r)\;;\;\hat{E}=(E-\epsilon_1)/
\sigma_1\\
\mbox{with}\;\;\epsilon_2 - \dis\frac{2\sigma_2}{\dis\sqrt{1-q}} \leq E_f 
\leq \epsilon_2 + \dis\frac{2\sigma_2}{\dis\sqrt{1-q}} \;\;
\mbox{and}\;\;\epsilon_1 - \dis\frac{2\sigma_1}{\dis\sqrt{1-q}} \leq E 
\leq \epsilon_1 + \dis\frac{2\sigma_1}{\dis\sqrt{1-q}} \;,\\
\\
\rho_{1:\co}(E) = \dis\frac{1}{\sigma_1}\;f_{qN}(\hat{E}|q)\;.
\earr \label{eq.npc2}
\ee
Note that the $q$ value for $\rho_{biv-\co}(E,E_f)$ and $\rho(E_f)$ need not be
same in general, i.e. $q \neq q^\prime$. Substituting all those in Eq.
(\ref{eq.npc2}) in Eq. (\ref{eq.npc1}) will give the following formula,
\be
\barr{l}
(\mbox{NPC})_E = \dis\frac{d}{3}\;\l[f_{qN}(\hat{E}|q)\r]^2\;
\l\{\dis\int_{y_0^i}^{y_0^f} dy\; \hat{\sigma}\;\dis\frac{\l[f_{biv-qN}(
\hat{E},y|\rho , q)\r]^2}{f_{qN}\l(\f{y-\hat{\Delta}}{\hat{\sigma}}
\l|q^\prime\r.\r)}\r\}^{-1}\;; \\
\\
\hat{\sigma} = \dis\frac{\sigma_f}{\sigma_2}\;,\;\;\;\hat{\Delta}=
\dis\frac{\epsilon_f -\epsilon_2}{\sigma_2}\;,\\
\\
y_0^i=max\l(\hat{\Delta}-\dis\frac{2\hat{\sigma}}{\dis\sqrt{1-q^\prime}}\;,\;
-\dis\frac{2}{\dis\sqrt{1-q}}\r)\;,\;\;
y_0^f=min\l(\hat{\Delta}+\dis\frac{2\hat{\sigma}}{\dis\sqrt{1-q^\prime}}\;,\;
\dis\frac{2}{\dis\sqrt{1-q}}\r)\;.
\earr \label{eq.npc3}
\ee
It is of interest in future to apply Eq. (\ref{eq.npc3}) to some realistic 
examples and also check in some examples if $\hat{\sigma} \sim 1$ and 
$\hat{\Delta} \sim 0$.

\begin{figure}
         \centering
         \includegraphics[width=6.5in,height=2.5in]{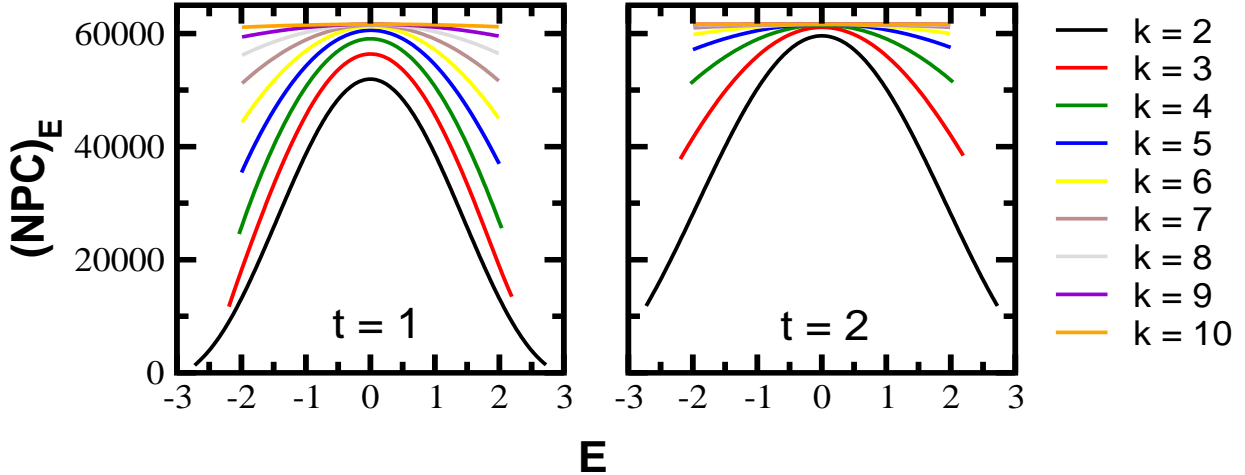}
     \caption{$(\mbox{NPC})_E $ given by Eq. \eqref{eq.npc3} as a function of 
     $E$ for various $k$ values with $t = 1$ (left panel) and $t = 2$ 
     (right panel). Results are shown for $m = 10$ fermions in $N = 20$ sp 
     levels and the parameters $q$ and $\rho$ are obtained using Eqs. 
     \eqref{eq.qbiv22} and \eqref{eq.qbiv23} respectively. We assume 
     $\hat{\sigma} = 1$, $\hat{\Delta} = 0$ and $q = q^\prime$. Note that $E$ in the figure is same as $\hat{E}$ in Eqs. (27) and (28).}
\label{fig:npc}
\end{figure}

Figure \ref{fig:npc} shows $(\mbox{NPC})_E $ given by Eq. \eqref{eq.npc3} as a
function of $E$ for various $k$ with $t = 1$ (left panel) and $t = 2$ (right
panel). Results are shown for $m = 10$ fermions in $N = 20$ sp levels and the
parameters $q$ and $\rho$ are obtained using Eqs. \eqref{eq.qbiv22} and
\eqref{eq.qbiv23} respectively; see Table \ref{tab3} for numerical values. We
assume $\hat{\sigma} = 1$, $\hat{\Delta} = 0$ and $q = q^\prime$. Note that the
$H$ matrix dimension for this system is $d=\binom{20}{10} = 184756$. It is seen
from the figure that for a given $t$, there is a transition from Gaussian form 
to the GOE result (GOE gives NPC to be $d/3 \sim 61585$) with
increasing $k$. This transition is faster for larger $t$.

\section{Conclusions}

Using lower order bivariate moments, it is established that the transition
strength densities generated by EGOE and EGUE random matrix ensembles follow
bivariate $q$-normal form. Formulas for the correlation coefficient $\rho^E$ and
the parameter $q^E$ are also given as a function of $(N,m,k,t)$ for $m$ fermions
in $N$ sp states with the Hamiltonian operator $H(k)$ and transition operator
$\co(t)$ represented by independent EGUE($k$) and EGUE($t$) respectively. These
formulas are expected to apply to EGOE and this follows from
\cite{kota,KoMa-15,MaKo-19}. In addition, application of the bivariate
$q$-normal to the NPC in transition strengths is described by deriving a formula
involving an integral.

Using $f_{biv-qN}$ and its extensions, it should be possible to address several
important issues in the subject of embedded ensembles with $k$-body interactions
[EE($k$)]. Some of these are as follows. (i) It is possible to study the measures
for wavefunction structure, as given by  the form of the strength functions
$F_k(E)$, number of principal components (NPC)$_E$ and information entropy
$S^{info}(E)$ \cite{kota,KS-01}, for a system of $m$ particles
(fermions or bosons) in a one-body mean-field with $N$ sp states and interacting
with a $k$-body force. Then, $H=h(1)+\lambda V(k)$ with $V(k)$ represented by
EGOE($k)$ or EGUE($k$). This is under investigation \cite{ChvMan}. Here, one
complication compared to the $k=2$ analysis given in \cite{KS-01} is that the
$V(k)$ with $k \geq 3$ will have more than two $U(N)$ tensorial parts with
tensorial rank $\nu=1,2,\ldots,k$ ($\nu=0$ part is not important here). (ii) It
may be possible to study the two-point function that gives the number variance
(fluctuations) for EGOE and EGUE using $q$-Hermite polynomials and the results
in Refs. \cite{Verb-2,LHKC}. (iii) Although $f_{CqN}(x|y; \rho , q)$ gives
$F_k(E)$ changing from Gaussian form to a semi-circle like form, this will not
give the Breit-Wigner (BW) form for $F_k(E)$ in any limit (BW form appears for
small values of $\lambda$). It is important to
study $q$ extended $t$-distribution so that the BW form is also included; see
\cite{kota} for the role of $t$-distribution in describing strength functions.
(iv) The bivariate $t$-distribution describes transition strength densities for
$\lambda$ small in $H=h(1)+ \lambda V(k)$ as shown in \cite{KCS-06} for $k=2$. Therefore,
it is important to study its $q$ extensions. (v) With other quantum numbers such
as  $J$ for the eigenstates, trivariate $q$-normal and in general multivariate
$q$-normal distributions may prove to be useful in random matrix theory with
$k$-body interactions; see \cite{Sza-1,Sza-2,Sza-4} for some properties of tri-
and multi-variate $q$-normal distributions. It is also of interest to
investigate the usefulness of the modified $q$-normal $\phi(x,t|q)f_{qN}(x|q)$
discussed in \cite{Sza-1}. 

\acknowledgments

Thanks are due to N. D. Chavda for useful discussions. M. V. acknowledges
financial support from UNAM/DGAPA/PAPIIT research grant IA101719.

\ed